\documentclass[prl,groupedaddress,showpacs,twocolumn]{revtex4}
\usepackage{graphicx}
\usepackage{amssymb}
\usepackage{amsmath}

\newcommand{\ped}[1]{\ensuremath{_{\rm #1}}}
\newcommand{\apex}[1]{\ensuremath{^{\rm #1}}}

\begin{document}
\title{Normal and superconducting properties of LiFeAs explained in the framework of four-band Eliashberg Theory}

\author{G.A. Ummarino}
\email{E-mail:giovanni.ummarino@infm.polito.it}
\author{Sara Galasso}
\author{D. Daghero}
\author{M. Tortello}
\author{R.S. Gonnelli}
 \affiliation{Istituto di Ingegneria e Fisica dei Materiali,
Dipartimento di Scienza Applicata e Tecnologia, Politecnico di
Torino, Corso Duca degli Abruzzi 24, 10129 Torino, Italy}
\author{ A. Sanna}
\affiliation{Max-Planck-Institut f\"ur Mikrostrukturphysik, Weinberg 2,
D-06120 Halle, Germany}

\begin{abstract}
In this paper we propose a model to reproduce superconductive and
normal properties of the iron pnictide LiFeAs in the framework of the
four-band $s\pm$ wave Eliashberg theory. A confirmation of
the multiband nature of the system rises from the experimental
measurements of the superconductive gaps and resistivity as function
of temperature. We found that the most plausible mechanism is the
antiferromagnetic spin fluctuation and the estimated values of the
total antiferromagnetic spin fluctuation coupling constant in the
superconductive and normal state are $\lambda\ped{tot}=2.00$ and
\mbox{$\lambda\ped{tot,tr}=0.77$}.
\end{abstract}


\maketitle
Recent ARPES measurements of iron superconductor LiFeAs report
four slightly anisotropic gaps \cite{GapUmez}. Their isotropic
values at 8~K are given by $\Delta\ped{1}=5.0$~meV,
$\Delta\ped{2}=2.6$~meV, $\Delta\ped{3}=3.6$~meV,
$\Delta\ped{4}=2.9$~meV and the critical temperature for this
compound is $T\ped{c}=18$~K \cite{Tapp}.
\begin{figure}
    \begin{center}
            \includegraphics[width=0.3\textwidth]{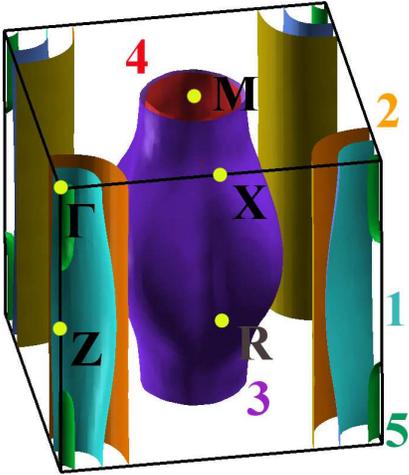}
    \end{center}
    \caption{Fermi surface of LiFeAs~\cite{computational}.}
    \label{Fig1}
\end{figure}

\begin{table}
\newcommand{\hs}{\hspace{0.2cm}}
\begin{tabular}{c|ccccc|c}
FS                      &   \hs 1       &   \hs2        &   \hs3        &   \hs4        &   \hs5        &   TOT \\
\hline
N(0)                        &   \hs 0.556   &   \hs 0.646   &   \hs 0.616   &   \hs 0.370   &   \hs 0.039   &   \hs 2.228 \\
$\omega_p^{\parallel ab}$       &   \hs  1.131  &   \hs  1.455  &   \hs  1.581  &   \hs  1.161  &   \hs  0.639  &   \hs  2.980\\
$\omega_p^{\parallel c}$    &   \hs  0.202  &   \hs  0.034  &   \hs  0.890  &   \hs  0.365  &   \hs  0.319  &   \hs  1.523\\
\end{tabular}
\caption{Fermi Surface resolved Kohn Sham properties: the Fermi
density of states $N(0)$  is given in states/spin/eV and plasma
frequencies $\omega_p$ in eV. $ab$ label the in-plane and $c$ for
the out-of-plane direction of the diagonals of the plasma tensor
\cite{PDdolghi}.}\label{tab:dft}
\end{table}

In an other work~\cite{Us} we disregarded the anisotropic part of
the gap values and we tried to reproduce the experimental data in
the framework of $s\pm$ wave multiband Eliashberg theory. At first,
we calculated~\cite{Us,computational,KS,PBE,espresso,PDdolghi} the
Fermi surface, depicted in FIG.~\ref{Fig1}: Five different sheets
are present, with two electron pockets centered near the M-point of
the Brillouin zone and three hole pockets around the $\Gamma$-point.
The 5-th sheet can be disregarded thanks to its low density of
states and size~\cite{Us} as can be seen in TABLE~\ref{tab:dft}. In
this way a four-band s-wave Eliashberg model~\cite{EE,ema} can be used
and eight coupled equations for the gaps $\Delta_{i}(i\omega_{n})$
and the renormalization functions $Z_{i}(i\omega_{n})$ have to be
solved. If $i$ is the band index (that ranges between $1$ and $4$) and
$\omega_{n}$ are the Matsubara frequencies, the imaginary-axis
equations are:
\begin{eqnarray}
\omega_{n}Z_{i}(i\omega_{n})&=&\omega_{n}+ \pi T\sum_{m,j}\Lambda^{Z}_{ij}(i\omega_{n},i\omega_{m})N^{Z}_{j}(i\omega_{m})+\nonumber\\
&&+\sum_{j}\big[\Gamma\ped{ij}+\Gamma^{M}\ped{ij}\big]N^{Z}_{j}(i\omega_{n})\,,
\label{eq:EE1}
\end{eqnarray}
\begin{eqnarray}
Z_{i}(i\omega_{n})\Delta_{i}(i\omega_{n})&=&\pi
T\sum_{m,j}\big[\Lambda^{\Delta}_{ij}(i\omega_{n},i\omega_{m})-\mu^{*}_{ij}(\omega_{c})\big]\times\nonumber\\
&& \times\Theta(\omega_{c}-|\omega_{m}|)N^{\Delta}_{j}(i\omega_{m}) \nonumber \\
&& +\sum_{j}[\Gamma\ped{ij}+\Gamma^{M}\ped{ij}]N^{\Delta}_{j}(i\omega_{n})\, ;
 \label{eq:EE2}
\end{eqnarray}
where $\Gamma\ped{ij}$ and $\Gamma^{M}\ped{ij}$ are the non magnetic
and magnetic impurity scattering rates, $\Theta(\omega_{c}-|\omega_{m}|)$ is the
Heaviside function and $\omega_{c}$ is a cutoff energy. Moreover, $\mu^{*}_{ij}(\omega\ped{c})$ are the elements of the $4\times 4$ Coulomb pseudopotential matrix and
$N^{\Delta}_{j}(i\omega_{m})=\Delta_{j}(i\omega_{m})/ {\sqrt{\omega^{2}_{m}+\Delta^{2}_{j}(i\omega_{m})}}$,
\mbox{$N^{Z}_{j}(i\omega_{m})=\omega_{m}/{\sqrt{\omega^{2}_{m}+\Delta^{2}_{j}(i\omega_{m})}}$.} Finally,
\[\Lambda^{Z}_{ij}(i\omega_{n},i\omega_{m})=\Lambda^{ph}_{ij}(i\omega_{n},i\omega_{m})+\Lambda^{sf}_{ij}(i\omega_{n},i\omega_{m})\]
\[\Lambda^{\Delta}_{ij}(i\omega_{n},i\omega_{m})=\Lambda^{ph}_{ij}(i\omega_{n},i\omega_{m})-\Lambda^{sf}_{ij}(i\omega_{n},i\omega_{m}).\]
Here the superscripts \emph{sf} and \emph{ph} mean ``antiferromagnetic spin fluctuations'' and
``phonons'', respectively. In particular,
\[\Lambda^{ph, sf}_{ij}(i\omega_{n},i\omega_{m})=2 \int_{0}^{+\infty}d\Omega \Omega
\frac{\alpha^{2}_{ij}F^{ph,sf}(\Omega)}{(\omega_{n}-\omega_{m})^{2}+\Omega^{2}},\]
and the electron-boson coupling constants are defined as
\begin{equation}
\lambda^{ph,sf}_{ij}=2\int_{0}^{+\infty}d\Omega\,\frac{\alpha^{2}_{ij}F^{ph,sf}(\Omega)}{\Omega}.
\end{equation}

The solution of eqs.\eqref{eq:EE1} and \eqref{eq:EE2} requires a
huge number of input parameters, then drastic approximations, compatible with the goal of
reproducing the essential physics of the problem, are
necessary to make the model solvable. As for many other
pnictides we assumed that~\cite{Us}: i) the total electron-phonon
coupling constant is small~\cite{Boeri}; ii) spin fluctuations
mainly provide interband coupling~\cite{Umma1}. This means that we
can set $\lambda^{ph}_{ii}=\lambda^{ph}_{ij}=0$,
$\mu^{*}_{ii}(\omega\ped{c})=\mu^{*}_{ij}(\omega\ped{c})=0$, i.e.
the electron-phonon coupling constant and the Coulomb
pseudopotential in first approximation compensate each other and
$\lambda^{sf}_{ii}=0$ (only interband SF coupling)~\cite{Umma1}.
However, within these assumptions, we were not able to reproduce the
observed gap values, and in particular the high value of
$\Delta_1$. In order to solve this problem it is necessary to
introduce an intraband coupling in the
first band ($\lambda_{11}\neq 0$).\\
The final matrix of the electron-boson coupling constants becomes
\begin{equation}
\lambda_{ij}= \left (
\begin{array}{cccc}
  \lambda_{11}                 &           0                   &               \lambda_{13}       &    \lambda_{14}  \\
  0                 &           0                   &              \lambda_{23}      &     \lambda_{24}   \\
  \lambda_{31}=\lambda_{13}\nu_{13}& \lambda_{32}= \lambda_{23}\nu_{23}   & 0  & 0\\
  \lambda_{41}=\lambda_{14}\nu_{14} & \lambda_{42}=\lambda_{24}\nu_{24}  & 0&  0\\
\end{array}
\right ) \label{eq:matrix}
\end{equation}
where $\nu_{ij}=N_{i}(0)/N_{j}(0)$ and $N_{i}(0)$ is the normal
density of states at the Fermi level for the $i$-th band ($i=1,
2,3,4$). We choose spectral functions with Lorentzian shape \cite{Umma1, Inosov} i.e:
\begin{equation}
\alpha_{ij}^2F_{ij}(\Omega)=
C_{ij}\big\{L(\Omega+\Omega_{ij},Y_{ij})-
L(\Omega-\Omega_{ij},Y_{ij})\big\}
\end{equation}
where $L(\Omega\pm\Omega_{ij},Y_{ij})=\frac{1}{(\Omega
\pm\Omega_{ij})^2+Y_{ij}^2}$ and $C_{ij}$ are normalization
constants, necessary to obtain the proper values of $\lambda_{ij}$
while $\Omega_{ij}$ and $Y_{ij}$ are the peak energies and
half-widths of the Lorentzian functions, respectively \cite{Umma1}.
In all the calculations we set
$\Omega_{ij}=\Omega_{ij}^{sf}=\Omega_0^{sf}=8$ meV \cite{Taylor},
and $Y_{ij}=Y_{ij}^{sf} = \Omega_{ij}^{sf}/2$ \cite{Inosov}. The
cut-off energy is $\omega_c = 18\,\Omega_0^{sf}$ and the maximum
quasiparticle energy is $\omega_{max}=21\,\Omega_0^{sf}$.
Bandstructure calculations (see TABLE \ref{tab:dft}) provide
information about the factors $\nu_{ij}$ that enter the definition
of $\lambda_{ij}$. In the end the model contains five free
parameters: The coupling constants
$\lambda_{13},\,\lambda_{23},\,\lambda_{14},\,\lambda_{24}$ and
$\lambda_{11}$. First of all we solved the imaginary-axis Eliashberg
equations \eqref{eq:EE1} and \eqref{eq:EE2} (actually we continued
them analytically on the real-axis by using the Pad\'e approximant
technique) and we fixed the free parameters in order to reproduce
the gap values at low temperature. The large number of free
parameters (five) may suggest that it is possible to find different
sets that produce the same results. On the contrary, as a matter of
fact, the predominantly interband character of the model drastically
reduce the number of possible choices. At this point there are no
more free parameters. We can calculate the critical temperature and
it turns out to be very close to experimental one ~\cite{Tapp}:
$T^{calc}_c=18.6$ K. In TABLE~\ref{tab:tab2} the obtained results
are summarized.
The problem of this model is the necessity of a so large intraband term $\lambda_{11}$
in order to give a physical interpretation of the experimental data~\cite{Us}.
\begin{table}
\begin{tabular}{c|c|ccccc|ccccc}
&$\lambda_{11}$    &$\lambda_{tot}$ &$\lambda_{13}$&$\lambda_{23}$&$\lambda_{14}$&$\lambda_{24}$&$\Delta_{1}$&$\Delta_{2}$&$\Delta_{3}$&$\Delta_{4}$&$T_{c}$\\
\hline
Exper. & -& -& -& -& -& -&5.0 & 2.6  & 3.6&  2.9&18\\
Theor. &0.0   &1.8 &  1.78  &  0.66 & 0.45  & 0.52  & 3.7 & 2.6  & 3.6&  2.9&  15.9\\
Theor. &2.1    & 2.0 & 1.15 & 0.80 &  0.45  & 0.30  &5.0 & 2.6  & 3.6&  2.9&  18.6\\
\end{tabular}
\caption{In the first row are shown the experimental data. The
second row concerns the pure intraband case ($\lambda_{ii}=0.0$)
while the third one concerns the case with a large intraband term
($\lambda_{11}=2.1$). The critical temperatures are given in K and
the gap values in meV. }\label{tab:tab2}
\end{table}\\

\begin{figure}
 \begin{center}
     \includegraphics[width=0.5\textwidth]{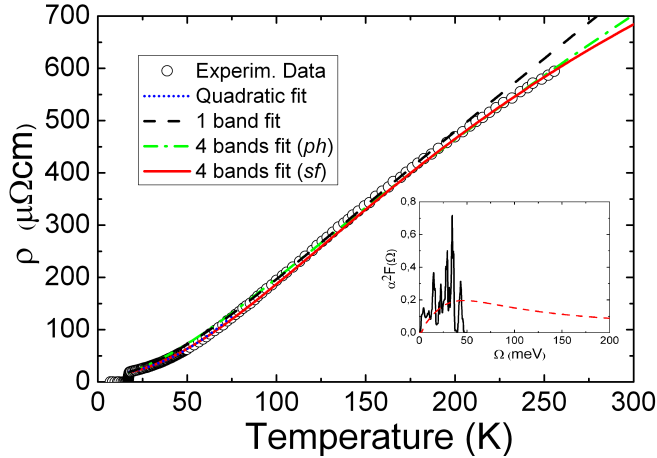}
  \end{center}
  \caption{Temperature dependence of resistivity in LiFeAs.
  Experimental data and calculated fits are reported.
  The black dashed line comes from a single-band model.
  Within a four-band model two different cases have been considered,
  one is obtained with the phononic spectra (green dash-dotted line) and one with the antiferromagnetic
  spin fluctuation spectra (red solid line). The inset shows the two normalized spectral function that have been used,
  the phonon spectra (black solid line) and the antiferromagnetic spin fluctuation spectra (red dashed line)}
  \label{Fig2}
  \end{figure}

Regarding the normal state~\cite{Heyer}, the resistivity saturation
at high temperature~\cite{Kasahara} suggests that the presence of
several sheets in the Fermi surface also affects the normal
state transport properties.\\
First of all we noticed (see  FIG.~\ref{Fig2}) that at low temperature \mbox{$\rho(T)\propto
T\apex{2}$} and this could
indicate that a non-phononic mechanism plays a relevant role
in the physics of this compound~\cite{Gur2D}.\\
To begin with, we tried to fit the data within a one-band
model~\cite{Allen, grim} (see eq.~\eqref{rho} with $i=1$) where the phonon
spectra has been taken from ref.~\cite{BorisPH} and the plasma
energy has been obtained by first principle calculation (see
TABLE~\ref{tab:dft}). The transport coupling constant and the value
of the impurities are considered as free parameters. The obtained
values are reported in TABLE~\ref{tab:tab3}, in particular
$\lambda\ped{tr,tot}=0.32$ which is in agreement with the calculated
value of the trasport electron-phonon coupling
constant~\cite{phonon}. However, as can be seen in FIG.~\ref{Fig2},
within a one-band model (black dashed line) the experimental data cannot be reproduced. \\
Phenomenological model~\cite{PhenModel} proposed to explain saturation at high temperature generally
assume the presence of parallel conductivity channels where one of them has a strong temperature
dependence and another one is characterized by a temperature-independent contribution.
\begin{table}
\newcommand{\hs}{\hspace{0.1cm}}
\begin{tabular}{c|cccccccc}
 &\hs $\lambda\ped{tr,tot}$&\hs $\lambda\ped{tr,3}$&\hs $\lambda\ped{tr,4}$&\hs $\gamma\ped{1}$&\hs $\gamma\ped{2}$&\hs $\gamma\ped{3}$&\hs $\gamma\ped{4}$&\hs$\Omega\ped{0}$ \\
 \hline
\emph{ph} 1 band & \hs 0.32 &\hs -&\hs -&\hs 0.90&\hs -&\hs -&\hs -&\hs -\\
\emph{ph} 4 bands & \hs 0.14 &\hs 0.44&\hs 0.10&\hs 5100&\hs 5100&\hs 0.65&\hs 550&\hs -\\
\emph{sf} 4 bands  & \hs 0.77&\hs 1.70&\hs 1.70&\hs 164&\hs 164&\hs
4.87&\hs 1.52&\hs 47
\end{tabular}
\caption{The first and second rows concern the phonon case while the
third one concerns the case of the antiferromagnetic spin
fluctuation spectral function. The $\gamma\ped{i}$ and
$\Omega\ped{0}$ are given in meV.} \label{tab:tab3}
\end{table}
\begin{figure}
 \begin{center}
     \includegraphics[width=0.5\textwidth]{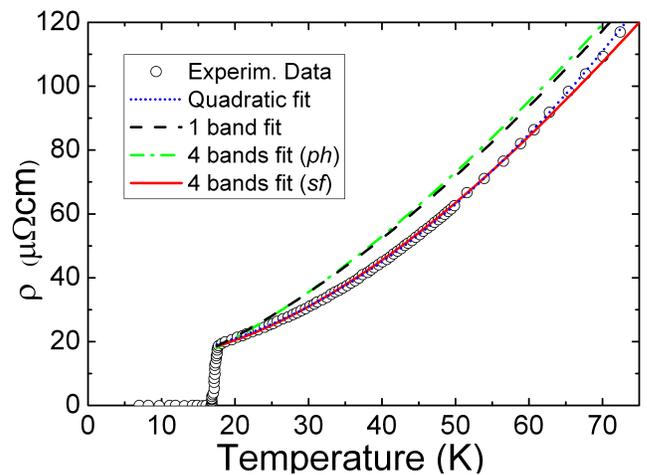}
  \end{center}
  \caption{Magnification of the previous figure. Resistivity at low temperature.}
  \label{Fig3}
  \end{figure}
In the wake of the model proposed for the superconducting state, we
propose a multiband model~\cite{MgB2,DolghiBaK} for analyzing the
resistivity data. We will examine two possible mechanism responsible
of resistivity: Phonons and antiferromagnetic spin fluctuations. The
theoretical expression of the resistivity as function of temperature~\cite{MgB2,DolghiBaK} is
given by the equation:
\begin{equation}
\frac{1}{\rho\ped{c}(T)}=\frac{\varepsilon\ped{0}}{\hbar}\sum_{i=1}^4\frac{(\hbar\omega\ped{pl,i})\apex2}{\gamma\ped{i}+W' \ped{i}(T)},
 \label{rho}
\end{equation}
where $\omega\ped{pl,i}$ is the bare plasma frequency of the $i$-band and
\begin{equation}
W' \ped{i}(T)=4\pi k\ped{B}T\int_0^\infty d\Omega
\left[\frac{\hbar\Omega/2k\ped{B}T}{\sinh\big(\hbar\Omega/2k\ped{B}T\big)}\right]\apex2 \frac{\alpha\ped{tr,i}\apex{2}F\ped{tr,i}(\Omega)}{\Omega},\\
 \label{W1}
\end{equation}
here $\gamma\ped{i}=\sum_{j=1}^4\Gamma\ped{ij}+\Gamma^{M}\ped{ij}$
is the sum of the inter- and intra-band non magnetic and magnetic
impurity scattering rates present in the Eliashberg equations and
\begin{equation}
\alpha\ped{tr,i}\apex{2}F\ped{tr,i}(\Omega)=\sum_{j=1}^4\alpha\ped{tr,ij}\apex{2}F\ped{tr,ij}(\Omega),
\end{equation}
where $\alpha\apex2\ped{tr}(\Omega)F\ped{tr,ij}(\Omega)$ are the transport spectral functions related to the Eliashberg functions \cite{Allen}. \\
If a normalized transport spectral function $\alpha\apex2\ped{tr}(\Omega)F'\ped{tr,i}(\Omega)$ is defined,
then $\alpha\apex2\ped{tr}(\Omega)F\ped{tr,ij}(\Omega)=\lambda\ped{tr,ij}\alpha\apex2\ped{tr}(\Omega)F'\ped{tr,ij}(\Omega)$ where the coupling constants are defined as for the standard Eliashberg functions.\\
In order to build a model as simple as possible, we chose all the
normalized transport spectral functions to be equal, then
$\alpha\apex2\ped{tr}(\Omega)F'\ped{tr,i}(\Omega)=\lambda\ped{tr,i}\alpha\apex2\ped{tr}(\Omega)F'\ped{tr}(\Omega)$
where
$\lambda\ped{i}=\sum_{j=1,..4} \lambda\ped{ij}$.\\
It has been shown that, at least for iron pnictides, this model can
have a theoretical support~\cite{DolghiBaK} depending on the
electronic structure of the compound. The basic idea, based on ARPES
and de Haas-van-Alphen data, is that the transport is drown mainly
by the electronic bands and that the hole bands have a weaker
mobility~\cite{dHvA}. Then the impurities are mostly present in the
hole bands and $\gamma\ped{1,2}\gg\gamma\ped{3,4}$, while the
transport coupling is much higher in bands 3 e 4 and this means that, at least as a
first approximation, $\lambda\ped{1}$ and $\lambda\ped{2}$ can be
fixed to be zero. In this way we will have two contributions almost
temperature independent and two which
change the slope of the resistivity with the temperature~\cite{DolghiBaK}. \\
Let us start with the phononic case. For simplicity we considered
all the spectral function to be proportional to the phonon spectra
used also in the previous fit~\cite{BorisPH}. As mentioned above the
transport spectral functions are similar to the standard Eliashberg
functions. The main difference is the behavior for
$\Omega\rightarrow 0$~\cite{Allen}, where the transport function
behaves like $\Omega^4$ instead of $\Omega^2$ as in the
superconducting state. So the condition
$\alpha\apex2\ped{tr}(\Omega)F\ped{tr}(\Omega) \propto \Omega\apex4$
has been imposed in the range $0<\Omega< k\ped{B}T\ped{D}/10$ and
then
\begin{eqnarray}
\alpha\apex2\ped{tr}(\Omega)F'\ped{tr}(\Omega)&=&b\ped{i}\Omega\apex{4}\vartheta(k\ped{B}T\ped{D}/10-\Omega)\nonumber\\
&&+ c\ped{i}\alpha\apex2\ped{tr}(\Omega)F''\ped{tr}(\Omega)\vartheta(\Omega-k\ped{B}T\ped{D}/10),\nonumber
\end{eqnarray}
where $T\ped{D}=240$ K is the Debye temperature~\cite{TD}, the
constant $b\ped{i}$ and $c\ped{i}$ are fixed by imposing the
continuity in $k\ped{B}T\ped{D}/10$, and the normalization to 1 and
$\alpha\apex2\ped{tr}(\Omega)F''\ped{tr}(\Omega)$ is proportional to
electron-phonon spectral function~\cite{BorisPH}.
$\alpha\apex2\ped{tr}(\Omega)F'\ped{tr}(\Omega)$
is shown in the inset of FIG.~\ref{Fig3}.\\
All the plasma frequencies are fixed by first principle calculations
(see TABLE I) and the coupling constants are considered as free
parameters as well as the impurities parameters. The best fit is
obtained with $\lambda\ped{tr,tot}=0.14$, as reported in
TABLE~\ref{tab:tab3}, which is in agreement with the hypothesis that
the phonon coupling in LiFeAs is very weak and the value of
$\lambda_{tr,4}$ almost does not influence the final result. However
the experimental data are not perfectly reproduced, as can be seen
by looking the green dash-dotted curve in FIG.~\ref{Fig3} and in
FIG.~\ref{Fig3}, moreover a huge quantity of impurity has been
necessary to obtain this theoretical curve and
this is not in agreement with the good quality of single crystal~\cite{Kasahara}.\\
Then we considered the case of antiferromagnetic spin fluctuations.
Now for $\Omega\rightarrow0$ the transport function behaves like
$\Omega^3$ instead of $\Omega$ as in the superconducting state. So
the condition
$\alpha\apex2\ped{tr}(\Omega)F\ped{tr}(\Omega)\propto\Omega\apex3$
has been imposed in the range $0<\Omega< \Omega\ped{0}/10$ and then
$\alpha\apex2\ped{tr}(\Omega)F'\ped{tr}(\Omega)=
b\ped{i}\Omega\apex{3}\vartheta(\Omega\ped{0}/10-\Omega)
+c\ped{i}\alpha\apex2\ped{tr}(\Omega)F''\ped{tr}(\Omega)\vartheta(\Omega-\Omega\ped{0}/10)$
and the constant $b\ped{i}$ and $c\ped{i}$ are fixed in the same way as before.\\
We choose as $\alpha\apex2\ped{tr}(\Omega)F''\ped{tr}(\Omega)$ the
theoretical antiferromagnetic spin fluctuation function in the
normal state \cite{Popovich}
\begin{equation}
\alpha\ped{tr}\apex{2}F''(\Omega)\propto\frac{\Omega\ped{0}\Omega}{\Omega\apex{2}+\Omega\ped{0}\apex{2}}\vartheta(\Omega-\Omega\ped{0}),
\end{equation}
where $\Omega\ped{0}$ is a free parameter: from the fit of experimental data we obtain $\Omega\ped{0}=47$ meV.\\
Also in this case the value of the free parameters are reported in
TABLE~\ref{tab:tab3} and FIG.~\ref{Fig2} depicts the obtained
results with the red solid line as well as the spectral function (in
the inset). The curve obtained by using the spin fluctuation spectra
better reproduce the experimental with a total coupling given by
$\lambda\ped{tr,tot}=0.77$ consistent with expectations, indeed is
smaller than the value in the superconducting state. Moreover the
parameters seems to better represent the LiFeAs sample: this is a
stoichiometric compound and the data have been taken from
measurements on a single
crystal sample, then the presence of a huge amount of impurities is not supported.\\
Of course we have done a draft simplification because the more plausible situation is the coexistence
of two mechanisms but certainly the antiferromagnetic spin fluctuactions constitute the main mechanism.
In conclusion we can say that in this compound the antiferromagnetic
spin fluctuations play an important role also in the normal state,
moreover information about the energy peak of the spectral function
and the total transport coupling constant have been extracted.



\end{document}